\documentclass[aps,prl,superscriptaddress,amsmath,amssymb,preprintnumbers
,twocolumn,showpacs]{revtex4}
\usepackage{amssymb}

\newcommand{\Ket}[1]{\left\vert#1\right\rangle}

\newcommand{\KetBra}[2]{\left\vert#1\right\rangle\left\langle#2\right\vert}
\newcommand{\Projector}[1]{\KetBra{#1}{#1}}
\newcommand{\MatrixEl}[3]{\left\langle#1\right\vert #2 \left\vert#3\right\rangle}

\newcommand{\ho}{\hat{H}_0}      
\newcommand{\freeze}{\Lambda}    
\newcommand{\hf}{\hat{H}_{12}}   
\newcommand{\heat}{\Gamma}       
\newcommand{\hh}{\hat{H}_{23}}   
\newcommand{\htot}
           {\hat{H}_{\freeze}}   
\newcommand{\htotzero}
           {\hat{H}_{\freeze=0}} 

\begin{document}


\title{Steering Distillation Processes through Zeno Effect}


\author{B. Militello}
\affiliation{INFM, MIUR and Dipartimento di Scienze Fisiche ed
Astronomiche dell'Universit\`{a} di Palermo, Via Archirafi, 36,
I-90123 Palermo, Italy}

\author{H. Nakazato}
\affiliation{Department of Physics, Waseda University, Tokyo
169-8555, Japan}

\author{A. Messina}
\affiliation{INFM, MIUR and Dipartimento di Scienze Fisiche ed
Astronomiche dell'Universit\`{a} di Palermo, Via Archirafi, 36,
I-90123 Palermo, Italy}


\begin{abstract}
A quantum system in interaction with a repeatedly measured one
undergoes a non-unitary time evolution pushing it into a subspace
substantially determined by the two-system coupling. The
possibility of suitably modifying such an interaction through the
generalized Quantum Zeno Effect addressing the system toward an
{\em a priori} decided target subspace is reported.
\end{abstract}


\maketitle


A distillation process, in its essence, is nothing but a
systematic way of driving a system toward a state nonorthogonal to
its initial condition. In some sense thus, it often represents the
realization of a projection operator. In recent years, efforts
have been made to realize distillation procedures since they might
be exploited to prepare and to control at one's will the state of
a quantum system. In the field of quantum technology\cite{qtech}
covering such research areas like quantum computation, quantum
information or quantum teleportation it is, for instance, usually
assumed that a desired state, which is necessary to start with a
specific quantum manipulation, can be in principle prepared. In
reality, however, it is not at all trivial and we have to specify
explicitly how such a preparation is actually implemented. Various
distillation processes have been proposed for this purpose so
far~\cite{qpp}\@.

Recently a general and new strategy of distillation specifically
addressed to bipartite systems has been
reported~\cite{NakazatoPRL}\@. Its goal is to drive the subsystem
of interest (called {\em Slave}) toward a final state by repeated
measurement acts performed on the auxiliary subsystem (called {\em
Master}), in interaction with the former, provided the latter part
is always found in the same state. This conditional procedure,
inspired by the Quantum Non-Demolition Measurements
(QND)~\cite{QND}\@, shares with the latter the difficulty of
singling out the suitable Master--Slave coupling allowing the
selection of a {\em prefixed} target state. Such an intrinsic
weakness poses limitations on the practical application of the
method.

In this letter we propose an effective way of overcoming this
structural drawback. Our scope is indeed to demonstrate the
systematic possibility of controlling the distillation mechanism,
furnishing conditions on the coupling to reach prefixed target
subspaces. The key tool to suitably modify the two-subsystem
interaction is the generalized Quantum Zeno Effect
(QZE)~\cite{FacchiPascazioPRL,QZE,FacchiPascazioReview}\@.


Consider two interacting systems, referred to as {\em Master}
($M$) and {\em Slave} ($S$). Let $\{\Ket{\phi_k}\}$ denote a basis
of $M$, and $\{\Ket{\varphi_{n}}\}$ indicate a basis of $S$. Let
us denote by $\hat{U}(\tau)=e^{-i\hat{H}\tau}$ the time evolution
operator ($\hbar=1$) for time $\tau$, $\hat{H}=\ho$ being the
compound system hamiltonian.

Perform a measurement of the Master state, and assume that the
result is $\Ket{\phi_0}$. Let the system evolve for time $\tau$
under $\hat H$, then perform another measurement of the Master,
and so on $N$ times. Assume that at each step the Master system is
found into the state $\Ket{\phi_0}$. Under such a hypothesis the
compound system is subjected to a non-unitary time evolution
described by
\begin{align}\label{NonUnitaryCompleteEvolutor}
  \nonumber
  \hat{W}^{(N)}(\tau)&\equiv
  \aleph_N\left[\Projector{\phi_0}\hat{U}(\tau)\right]^N
  \Projector{\phi_0}\\
  &=\aleph_N\Projector{\phi_0}\left[\hat{V}(\tau)\right]^{N},
\end{align}
where
\begin{equation}\label{NonUnitaryOneStepEvolutor}
  \hat{V}(\tau)\equiv
  \MatrixEl
  {\phi_0}
  {e^{-i\hat{H}\tau}}
  {\phi_0}
\end{equation}
is a non-unitary operator describing the transformation that the
Slave state undergoes owing to both the time evolution
$\hat{U}(\tau)$ and the projection following the successful
measurement act on $M$\@. The normalization constant $\aleph_N$
takes into account the probability of finding $M$ in the state
$\Ket{\phi_0}$ $N$ times.

Let us solve the eigenvalue problem related to $\hat{V}(\tau)$.
Since this operator is not hermitian, generally its right and left
eigenvalue problems turn out to have different solutions. Assume
now that it is possible to {\em diagonalize} the operator
$\hat{V}(\tau)$ in the standard way~\cite{NUY04}
\begin{equation}\label{NonUnitaryEvolutorDecomposition}
  \hat{V}(\tau)=\sum_{k} \gamma_k \hat{P}_{k},
\end{equation}
where $\{\hat{P}_{k}\}$ are orthogonal projection operators satisfying
the completeness relation $\sum_{k}\hat{P}_{k}=\hat{1}_{S}$,
$\hat{1}_{S}$ being the Slave identity operator.

Introducing $\gamma=\max_{k}\{|\gamma_k|\}$, for large enough $N$,
one finds
\begin{equation}\label{ReductionToProjector}
  \hat{W}^{(N)}(\tau)\approx
  \aleph_N\Projector{\phi_0}\sum_{k:|\gamma_k|=\gamma} \gamma_k^N \hat{P}_{k}
\end{equation}
If a unique eigenvalue of $\hat{V}(\tau)$ having modulus $\gamma$
exists and in addition it is non-degenerate, then the
distillation procedure realizes just a \lq single-state selection.\rq

The success of such a procedure is based upon the realization of
some specific conditions. First of all, observe that what we have
done is implementing a projection operator into a Slave subspace
determined by the specific Master-Slave interaction and by the
repeatedly detected Master state. Hence in order to reach the goal
of distilling a state, such a state should be {\em included} in
the initial state. Moreover, the evolution leading to the
target state is not deterministically followed by the system,
instead it is a \lq conditional\rq\ one. In fact there are $N$
{\em stochastic} intermediate steps (the measurement acts), and each
one of them should be done successfully, that is, at each Master
measurement step the result should be $\Ket{\phi_0}$ (or some
other prefixed Master State, but {\em always the same}).

Assume that the {\it orthonormal\/} basis $\{\Ket{\varphi_{n}}\}$
constitutes the eigenstates of $\hat{V}(\tau)$. Accordingly, the
action of the unitary operator $\hat{U}(\tau)$ on the state
$\Ket{\varphi_{n}}\otimes\Ket{\phi_0}$ induces transitions to
other Master states but leaves unchanged the Slave state which
factorizes $\Ket{\phi_0}$, i.e.,
$\Ket{\varphi_{n}}\otimes\Ket{\phi_0}$ $\longrightarrow$
$\sqrt{\wp_n}(\tau)e^{i\xi_n(\tau)}\Ket{\varphi_{n}}\otimes\Ket{\phi_0}+$
$\sum_{j\not=0}\sum_{m}c_{j,m}(\tau)\Ket{\varphi_m}\otimes\Ket{\phi_j}$,
$\wp_n(\tau)$
 being the relevant survival probability. Therefore
one immediately finds
\begin{equation}\label{Decompositon_SurvProb}
\hat{V}(\tau)=\sum_{n}\sqrt{\wp_n(\tau)}e^{i\xi_n(\tau)}
\Projector{\varphi_n}.
\end{equation}
Such a decomposition gives a precise physical meaning to the
moduli of the eigenvalues of $\hat{V}(\tau)$ and shows that the
distilled (preserved) Slave states (corresponding to higher
eigenvalue moduli) are those undergoing a unitary evolutions
(between two measurement acts) which do not induce (or induce
smaller) abandon of the initial Master states $\Ket{\phi_0}$ in
favor of the others $\Ket{\phi_j}$, $j\not=0$. Therefore, each
state $\Ket{\varphi_n}$ may be thought of as a \lq channel\rq\ of
probability loss, which may be opened or closed depending on the
specific features of $\hat{U}(\tau)$.

Consider now the case wherein all the transition channels are
going to be closed, i.e. all transitions from $\Ket{\phi_0}$ to
different Master states are hindered. Such a circumstance may
occur under the presence of an {\em external agent}, the {\em
freezing agent}, related to a hamiltonian $\hf$ responsible for an
additional coupling between Master and Slave. Hence, in the lapse
between two Master detections, the dynamics of the compound system
is governed by $\htot=\ho+\freeze\hf$, where $\freeze$ is the
strength of the new coupling term. Let us denote by
$\hat{U}_{\freeze}(\tau)\equiv e^{-i\htot t}$ the relevant
evolution operator.

It is well known that the stronger the new interaction term (i.e.,
the bigger $\freeze$) is, the more the dynamics is governed by
$\hf$~\cite{FacchiPascazioPRL}\@. This means that the time
evolution exhibits invariant subspaces related substantially to
$\ho$ when $\freeze$ approaches $0$ and those related to the
structure of $\hf$ as $\freeze$ grows
enough~\cite{FacchiPascazioPRL}\@. Depending on the specific
features of $\ho$ and $\hf$, it is possible to provide an
interpretation of this change of dynamical regime in terms of the
{\em Continuous Measurement Quantum Zeno Effect} or {\em Zeno
Dynamics}~\cite{FacchiPascazioReview}\@. To better understand this
point of view, consider the particular case wherein the
hamiltonian $\ho$ governs Slave-state-dependent oscillations
between the two Master states $\Ket{\phi_0}$ and $\Ket{\phi_1}$,
while $\hf$ couples states characterized by the Master states
$\Ket{\phi_1}$ and $\Ket{\phi_2}$. In this case the second
coupling term may be thought of as an {\em observer} that,
watching at the system following the unperturbed dynamics governed
by $\ho$, is able to freeze it into the initial Master state
$\Ket{\phi_0}$~\cite{FacchiPascazioReview}\@. The larger the \lq
watching-strength\rq\ ($\freeze$) is, the stronger the \lq
freezing\rq\ effect is. Since $\hf$ is, generally speaking,
responsible for a Slave-state-dependent quantum Zeno effect, the
distillation may be made ineffective/effective, depending on the
corresponding Hilbert subspaces of the Slave.

The ability of controlling the \lq channel opening/closing\rq\ is
improvable when the more general case of {\em Hierarchically
Controlled Dynamics}~\cite{MilitelloFortPh,FacchiPascazioPRL} is
considered, instead of the Zeno effect. In fact, the action of a
subsequent coupling, let us denote it by $\hh$ assuming for
simplicity that it involves Master states $\Ket{\phi_2}$ and
$\Ket{\phi_3}$ only, can hinder the hindering effect of $\hf$\@.
As a very simplified toy-model clarifying this point, let us
consider, parenthetically, the following $4\times 4$ tridiagonal
matrix, expressing a coupling scheme between the states of a
four-level system:
\begin{equation}\label{Tridiagonal_4by4_Matrix_OnlyMaster}
\Xi=\left(
\begin{array}{cccc}
0&\Omega&0&0\\
\Omega^*&0&\freeze&0\\
0&\freeze^{*}&0&\heat\\
0&0&\heat^*&0\\
\end{array}
\right).
\end{equation}
The four-level system, whose relevant dynamics is here described,
performs Rabi oscillations between the two lowest states (those
coupled by $\Omega$) when $\freeze=\heat=0$. Such a dynamical
regime corresponds to the unperturbed time evolution. As $\freeze$
is made non-vanishing, the dynamics becomes more and more
complicated until the condition $\freeze\gg\Omega$ is reached. In
such a situation, the dynamics of the lowest level is frozen, and
this is just the continuous measurement quantum Zeno effect
already recalled. Once $\heat$ is also adjusted as a non-vanishing
coupling constant, an unexpected phenomenon does happen: as
$\heat$ grows up, the effect due to the strong coupling $\freeze$
between the second and the third levels becomes weaker and weaker,
up to the point, identified by the condition $\heat\gg\freeze$,
wherein the original Rabi oscillations are completely restored. In
agreement with
Refs.~\cite{FacchiPascazioPRL,FacchiPascazioReview}, one can
interpret $\freeze$- and $\heat$-couplings as \lq\lq
continuous\rq\rq\  measurements so that it is possible to give the
following {\em metaphorical} statement: \lq\lq a watched pot never
boils\rq\rq ($\freeze\gg\Omega$, $\heat=0$) but \lq\lq a watched
cook can freely watch a boiling pot\rq\rq($\freeze\gg\Omega$,
$\heat\gg\freeze$)~\cite{FacchiPascazioPRL}.

In passing, we mention the fact that the {\em hierarchical chain}
of interactions in principle may be {\em extended} (with further
rings), maintaining the same substantial features: the last ring
of the chain is able to destroy the effects of the previous one
depending on how the relevant coupling strengths compare to each
other. For instance we could add a fifth level and a fourth
coupling involving the fourth and fifth levels. In this way, we
obtain a $5\times 5$ tridiagonal matrix. In concomitance with an
increase of the fourth-coupling strength, the hindering of
inhibition given by the third coupling ($\heat$) is hindered.

Coming back to our original problem, since the third coupling,
$\hh$, is Slave-state-dependent too, it can differently exert its
effect subspace by subspace. Such a circumstance, of course,
enlarges our ability of controlling the dynamical behavior of the
compound system and hence the distillation process.

In the light of such considerations, we can express the main
result of this paper as follows. A continuous measurement quantum
Zeno effect, involving Master states in a Slave-state-dependent
way, is able to hinder the $\ho$-governed unperturbed dynamics,
eventually closing channels for Slave-state probability loss,
which are open in the unperturbed dynamics. Hence the final
distillation result may be predicted and, moreover, {\em a priori
decided}, by exploiting the very clear mechanism of
channel-closing or opening in connection with our will of {\em
preserving} or {\em non-preserving} the corresponding Slave
states. Moreover, we can go through a hierarchically controlled
dynamics of the compound Master-Slave system, improving our
ability in implementing the filter.

Let us now consider, as an example of this strategy, a three-level
system (the Master) coupled to a harmonic oscillator (the Slave).
Such a physical system may be realized in the context of trapped
ions~\cite{ManiscalcoFortPh}.

As is well known, a time-dependent quadrupolar electric field is
able to confine a charged particle, providing an effective
quadratic potential that induces a harmonic motion. When the
confined particle is an ion, the complete result is a compound
system possessing both fermionic and bosonic degrees of freedom,
the first ones describing the internal motion of the electrons
with respect to the atomic nucleus, the second ones describing the
ion center-of-mass motion. In most of the experiments, only a few
atomic states are really involved in the dynamics and a single
vibrational mode is considered. Following
Ref.~\cite{ManiscalcoFortPh}, it is possible to realize an
experimental setup which involves, in the dynamics, only the
following three atomic levels,
$\Ket{g}:=\Ket{^{2}S_{1/2},F=1,m_F=1}$,
$\Ket{e_1}:=\Ket{^{2}S_{1/2},F=2,m_F=2}$
$\Ket{e_2}:=\Ket{^{2}S_{1/2},F=2,m_F=1}$, using a magnetic field
of 1 mT to obtain the useful level splittings
($\omega(e_2)-\omega(e_1)\approx 100$ MHz,
$\omega(g)-\omega(e_1)\approx 1$ GHz), and exploiting the
auxiliary level $\Ket{^{2}P_{1/2},F=2,m_F=2}$ to realize Raman
coupling schemes.

Consider now the action of two effective (i.e., implemented via
Raman schemes) lasers, one tuned to the $p$th blue or red sideband
of the atomic transition $\Ket{g}\rightarrow\Ket{e_1}$, and the
other to the $q$th blue or red sideband of the atomic transition
$\Ket{e_1}$ $\rightarrow$ $\Ket{e_2}$. The relevant
interaction-picture hamiltonian in the {\em Rotating Wave
Approximation} is given by\cite{MilitelloPLA}
\begin{align}\label{Model_3lev_Oscillator}
  \nonumber
  \htot=&\Omega\left[f_p(\hat{a}^{\dag}\hat{a}, \eta_1)\hat{a}^p\KetBra{e_1}{g}\;+\;{\rm H.c.}\;\right]\\
        &+\freeze\left[f_q(\hat{a}^{\dag}\hat{a}, \eta_2)\hat{a}^q\KetBra{e_2}{e_1}\;+\;{\rm H.c.}\;\right],
\end{align}
where $p$ and $q$ are integer numbers related to the specific
choice of the laser frequencies (sidebands), $\Omega$ and
$\freeze$ are coupling constants proportional to the laser
intensities, $\eta_j$ ($j=1,2$) are the relevant Lamb--Dicke
parameters expressing the ratio between the vibrational ground
state oscillation amplitudes and the laser wavelengths, $\hat{a}$
is the harmonic-oscillator annihilation operator (here the
notation is such that $\hat{a}^{-s}\equiv\hat{a}^{\dag s}$ for
positive $s$), while $\Ket{g}$, $\Ket{e_1}$ and $\Ket{e_2}$ are
the three atomic states effectively involved in the dynamics. The
functions $f_p$ and $f_q$ express nonlinear vibrational energy
dependence of vibronic couplings and are such that for very small
Lamb--Dicke parameters ($\eta\ll 1$), they almost approach unity,
while for larger values of $\eta$ they exhibit a strongly
nonlinear behavior in the {\em variable} $\hat{a}^{\dag}\hat{a}$,
and possesses some zeroes.

Considering the unitary evolution due to $\htot$ in
\eqref{Model_3lev_Oscillator} in the special regime $\freeze=0$
and $p=0$, and repeatedly detecting the atomic state $\Ket{g}$
leads to the standard quantum non-demolition measurements, by
which it is possible to extract, i.e. distill, a number
state~\cite{QND,NakazatoPRL}. More in detail, denoting by
$\Ket{n}$ the harmonic oscillator Fock states, the effective
non-unitary evolution operator acting upon the vibrational state
is given by
\begin{align}\label{NonUnitaryEvolutorQND}
  \nonumber
  \hat{V}_{\freeze=0}(\tau)\equiv&\MatrixEl{g}{e^{-i\htotzero\tau}}{g}
  =\cos\left[\Omega f_0(\hat{a}^{\dag}\hat{a}, \eta_1)\tau\right]\\
  =& \sum_n \cos\left[\Omega f_0(n, \eta_1)\tau\right]\Projector{n}.
\end{align}
Since in this case the Fock states are eigenstates of
$\hat{V}(\tau)$, we have $\Ket{\varphi_n}:=\Ket{n}$. It is
possible to choose $\tau=\tau_{\overline{n}}$ such that $\Omega
f_0(\overline{n},\eta_1)\tau_{\overline{n}}=\pi$. Taking into
account that for $n\not=\overline{n}$, $f_0(n,\eta_1)$ and
$f_0(\overline{n},\eta_1)$ are incommensurable, the eigenvector
$\Ket{\overline{n}}\otimes\Ket{g}$, and {\em it only}, is preserved by
the distillation procedure ($|\cos\left[\Omega
f_0(\overline{n},\eta_1)\tau\right]|=1$), while all the others are
partially destroyed at each atomic state detection 
($|\cos\left[\Omega f_0(n,\eta_1)\tau\right]|<1$ for
$n\not=\overline{n}$).

Consider now the effect of the freezing agent, i.e., assume
$\freeze\not=0$. In this case the total Hamiltonian, $\htot$, is
substantially characterized by three-dimensional invariant
subspaces, $\{\Ket{n}\otimes\Ket{g}, \Ket{n-p}\otimes\Ket{e_1},
\Ket{n-p-q}\otimes\Ket{e_2}\}$, wherein the operator may be represented
as a $3\times 3$ block of the form
\begin{equation}\label{TridiagonalMatrix}
\Xi_n=\left(
\begin{array}{ccc}
0&\Omega_{n}&0\\
\Omega_{n}^*&0&\freeze_{n}\\
0&\freeze_{n}^{*}&0
\end{array}
\right),
\end{equation}
with $\Omega_n=$ $\Omega f_p(n-p,\eta_1)\sqrt{{n!}/{(n-p)!}}$,
$\freeze_n=$ $\freeze
f_q(n-p-q,\eta_2)\sqrt{{(n-p)!}/{(n-p-q)!}}$\;. Of course,
depending on $p$ and $q$, there could exist also invariant
doublets and singlets. For instance, in correspondence to $n-p<0$
there is a singlet, while if $n-p\ge 0$ and $n-p-q<0$ there is a
doublet.

It is straightforward to evaluate the non-unitary operator
$\hat{V}_{\freeze}\equiv\MatrixEl{g}{\hat{U}_{\freeze}(\tau)}{g}$.
In particular, in the case $p=0$ (previously analyzed in absence of
freezing agent), we obtain
\begin{align}
  \nonumber
  \hat{V}_{\freeze}(\tau)
     =&\sum_{n=0}^{q-1} \cos\left[\Omega
     f_0(n,\eta_1)\tau\right]\Projector{n}\\
     &+\sum_{n=q}^{\infty}
    \frac{|\freeze_{n}|^2+|\Omega_{n}|^2\cos\left(\omega_{n}\tau\right)}
         {|\freeze_{n}|^2+|\Omega_{n}|^2}\Projector{n}
\end{align}
with $\omega_n=\sqrt{|\freeze_{n}|^2+|\Omega_{n}|^2}$. The
operator is diagonal in the Fock basis.

It is easy to see that in correspondence to continuous measurement
quantum Zeno effect ($\freeze_n\gg\Omega_n$) in the $n$th subspace
(the subspace the state $\Ket{n}\otimes\Ket{g}$ belongs to), the
channel of probability loss is closed, otherwise it could be open.
Therefore, the ability of controlling the dependence on $n$ of the
quantum Zeno effect incoming, plays the role of a real {\em handle
grip} by which it is possible to open and close distillation
channels. As a specific application, consider the dynamical regime
characterized by $\freeze\gg\Omega$, $p=0$ and, in order to have
$f_q(n-q,\eta_2)$ almost constant and non-vanishing for all Fock
states, $\eta_2\ll 1$. Under such hypotheses,
$\freeze_n\gg\Omega_n$, $\forall\;n$. If the measurement interval
$\tau$ is not fine-tuned, i.e., $\tau\not=\tau^{(k)}_0,
\tau^{(k)}_1,\ldots,\tau^{(k)}_{q-1}$, $\forall\; {\rm integer
}\;k$, where $\tau^{(k)}_j$ satisfies $\Omega
f_0(j,\eta_1)\tau^{(k)}_j=k\pi$, all Fock states with an
excitation number less than $q$ are eliminated in the course of
repeated measurements, hence realizing the projector
$\hat{1}-\sum_{n=0,1,\ldots,q-1}\Projector{n}$. That is,
\begin{equation}
\left[\hat{V}_{\freeze}(\tau)\right]^N \xrightarrow{\text{large
}N} \;\;\hat{1}-\sum_{n=0,1,\ldots,q-1}\Projector{n}.
\end{equation}

As another specific application, consider the implementation of
the projector $\hat{1}-\Projector{\overline{n}}$. In order to
reach the goal, we consider the case where $\freeze\gg\Omega$,
$p=q=0$, $\eta_2$ large enough to make the zeroes of $f_0$ {\em
visible} also for not extremely high $\overline n$:
$\;\freeze_{\overline{n}}=$ $\freeze f_0(\overline{n},\eta_2)=0$,
and $\tau$ is such that
$\cos(\omega_{\overline n}\tau)\not=1$. In this case, all channels
result to be closed (because of the condition
$\freeze_n\gg\Omega_n$, for $n\not=\overline{n}$) except for that
related to the $\overline{n}$th subspace (being
$\freeze_{\overline{n}}=0$).


In conclusion, in this paper we have addressed the general problem
of how to project a quantum system in a desired {\em prefixed}
subspace. We have indeed improved the distillation approach based
upon repeated measurements on a part of a bipartite
system\cite{NakazatoPRL}\@ introducing the generalized Quantum
Zeno Effect as a control mechanism of the
two--subsystem--interaction determining the target subspace. In
other words, our main and novel result is that the presence of
such a control mechanism acts upon the interaction transforming
the original distillation process into a driven distillation. We
conclude emphasizing that the method and the idea presented in
this letter are general and then exploitable in different physical
context both for fundamental and technological scopes.


This work is partly supported by the bilateral Italian-Japanese
project 15C1 on \lq\lq Quantum Information and Computation\rq\rq\
of the Italian Ministry for Foreign Affairs, by a Grant for The 21st
Century COE Program (Physics of Self-Organization Systems) at Waseda
University and a Grant-in-Aid for Priority Areas Research (B)
(No.~13135221), both from the Ministry of Education, Culture, Sports,
Science and Technology, Japan, and by a Grant-in-Aid for Scientific
Research (C) (No.~14540280) from the Japan Society for the Promotion
of Science.

\end{document}